\documentclass[aps,prd,nofootinbib,showpacs,superscriptaddress,12pt]{revtex4}
\usepackage{txfonts}
\usepackage{graphicx}
\usepackage{dcolumn}
\usepackage{bm}
\usepackage{amssymb}
\usepackage{latexsym}

\begin{document}

\title{\bf Analysis on a General Class of Holographic Type Dark Energy Models}
\author{Zhuo-Peng~Huang}
\email[Electronic address: ]{zphuang@nudt.edu.cn} \affiliation{Department of Physics, National University of Defense Technology, Hunan 410073, China}
\affiliation{State Key Laboratory of Theoretical Physics (SKLTP)\\
Kavli Institute for Theoretical Physics China (KITPC) \\
Institute of Theoretical Physics, Chinese Academy of Sciences, Beijing 100190, China}
\author{Yue-Liang~Wu}
\email[Electronic address: ]{ylwu@itp.ac.cn}
\affiliation{State Key Laboratory of Theoretical Physics (SKLTP)\\
Kavli Institute for Theoretical Physics China (KITPC) \\
Institute of Theoretical Physics, Chinese Academy of Sciences, Beijing 100190, China}

\date{\today}

\begin{abstract}

We present a detail analysis on a general class of holographic type dark energy models characterized by the length scale $L=\frac1{a^n(t)}\int_0^t dt'~a^m(t')$. We show that $n \geq 0$ is required by the recent cosmic accelerated expansion of universe. In the early universe dominated by the constituent with constant equation of state $w_m$, we have $w_{de}\simeq -1-\frac{2n}{3}$ for $n \geq 0$ and $m<0$, and $w_{de}\simeq-\frac23(n-m)+w_m$ for $n > m \geq 0$. The models with $n > m \geq 0$ become single-parameter models like the $\Lambda$CDM model due to the analytic feature $\Omega_{de}\simeq \frac{d^2}4(2m+3w_m+3)^2a^{2(n-m)}$ at radiation- and matter-dominated epoch. Whereas the cases $n=m\geq 0$ should be abandoned as the dark energy cannot dominate the universe forever and there might be too large fraction of dark energy in early universe, and the cases $m> n \geq 0$ are forbidden by the self-consistent requirement $\Omega_{de}\ll1 $ in the early universe. Thus a detailed study on the single-parameter models corresponding to cases $n >m \geq 0$ is carried out by using recent observations. The best-fit analysis indicates that the conformal-age-like models with $n=m+1$, i.e. $L\propto\frac1{Ha}$ in early universe, are more favored and also the models with smaller $n$ for the given $n-m$ are found to fit the observations better. The equation of state of the dark energy in models with $n=m+1 >0$ transits from
$w_{de}<-1$ during inflation to $w_{de}>-1$ in radiation- and matter-dominated epoch, and then back to $w_{de}<-1$ eventually. The best-fit result of the case $(n=0, m=-1)$ which is so-called $\eta$HDE model proposed in \cite{Huang:2012xm} is the most favorable model and compatible with the $\Lambda$CDM model.
\end{abstract}

\maketitle


\section{Introduction}

Since the discovery of the current accelerated expansion of the universe \cite{Riess:1998cb, Perlmutter:1998np}, many efforts have been made to explain the physical
essence behind this phenomenon of acceleration \cite{SS2000,Peebles:2002gy, Padmanabhan:2002ji, Copeland:2006wr, Tsujikawa:2010sc, Li:2011sd}. Within the framework of the
general relativity, the acceleration can be attributed to the existence of a mysterious negative pressure component named phenomenally as dark energy. A positive
cosmological constant, which fits to the observations well, is currently still the simplest candidate for dark energy. However, it is plagued with the fine-tuning
problem and the coincidence problem \cite{Weinberg:1988cp}. According to the holographic principle \cite{'tHooft:1993gx, Susskind:1994vu} that the number of degrees
of freedom of a physical system scales with the area of its boundary rather than its volume. This might indicate that degrees of freedom are overcounted in the
calculation of the vacuum energy in effective field theory. The author in \cite{Cohen:1998zx} suggested that the ultraviolet (UV) cutoff $\Lambda_{\rm uv}$ in the
effective field theory should be related to the infrared (IR) cutoff $L$ due to the limit set by the formation of a black hole, i.e., in terms of the natural units,
 \begin{equation}
    L^3\Lambda_{\rm uv}^4 \lesssim LM_p^2  ~,
 \end{equation}
where $M_p^2=1/(8\pi G)$ with $G$ the Newton's~constant. Such relation leads to a dramatic depletion of quantum states, which results in a much small vacuum energy density,
 \begin{equation}
    \rho_{\rm vac} \sim \Lambda_{\rm uv}^4 \sim  M_p^2L^{-2}  ~ \label{vac}.
 \end{equation}
When the IR cutoff $L$ is comparable to the current Hubble radius, the fine-tuning problem of the cosmological constant would not exist \cite{Cohen:1998zx}. Some
interesting studies on cosmology with holography have been carried out \cite{Horava:2000tb,Thomas:2002pq,Fischler:1998st,Bousso:1999xy}. In ref.\cite{Huang:2012xm},
we have shown that when the IR cutoff $L$ is characterized by the total comoving horizon of the universe, the inflation \cite{Guth:1980zm, S1980,Linde:1981mu,
Albrecht:1982wi} in early universe makes the vacuum energy given in Eq.(\ref{vac}) behave as a cosmological constant without the fine-tuning problem as well as the
coincidence problem, which provided an interesting insight on the cosmological constant as the simplest and best candidate for the dark energy. In alternative, we
have found in \cite{Huang:2012nz,Huang:2012gd} that when the IR cutoff $L$ is taken to be a conformal-age-like length, the dark energy density given by the vacuum
energy in Eq.(\ref{vac}) is almost insensitive to the inflation in early universe.

In the spirit of the holographic principle \cite{'tHooft:1993gx, Susskind:1994vu, Cohen:1998zx}, some interesting holographic dark energy models in which the dark
energy density is assumed to scale as $\rho_{de}\propto M_P^2L^{-2}$ with $L$ being some characteristic length scale of the universe have been proposed and studied
by many groups \cite{Li:2004rb,Hsu:2004ri,Huang:2004ai,Gong:2004fq,Myung:2004ch,Pavon:2005yx,
Wang:2005jx,Nojiri:2005pu,Hu:2006ar,Chen:2006qy,Li:2006ci,Setare:2006wh,Cai:2007us,Wei:2007ty,Gao:2007ep,Granda:2008dk,Granda:2008tm,Gong:2009dc,
Jamil:2009sq,Feng:2009hr,delCampo:2011jp,Gao:2011,Zhang:2012qr,Wei:2012wt,lingyi}. Particularly, the first holographic dark energy model (HDE) was proposed in
\cite{Li:2004rb} by identifying $L$ with the radius of the future event horizon, i.e. $a(t)\int_{t}^{\infty}\frac{dt'}{a(t')}$,  here $a(t)$ is the scale factor of
the universe at cosmic time $t$. Alternatively, the total comoving horizon $\int_{0}^{t}\frac{dt'}{a(t')}$ including the very large primordial part generated by
inflation was taken in \cite{Huang:2012xm} as the IR cutoff to establish the so-called $\eta$HDE model, as a consequence, the holographic dark energy behaves almost
like a cosmological constant. On the other hand, a conformal-age-like length $\frac{1}{a^4(t)}\int_0^tdt'~a^3(t') $ motivated from the four dimensional spacetime
volume at cosmic time $t$ of the flat FRW universe was adopted in \cite{Huang:2012nz} to build the holographic type dark energy model(CHDE). Based on different physical
origin, the age $\int_{0}^{t}dt'$ of the universe is chosen as the IR cutoff to construct the agegraphic dark energy model(ADE) in \cite{Cai:2007us}, its improved
new version(NADE) was proposed with replacing the age of the universe by the conformal age $\int^{t}\frac{dt'}{a(t')}$ of the universe \cite{Wei:2007ty}.

From the above considerations, it is interesting to investigate a general class of holographic type dark energy models with the characteristic length scale given by the
following form
\begin{equation}
    L=\frac1{a^n(t)}\int_0^t dt'~a^m(t') ~. \label{L0}
\end{equation}
Here we will focus on the case with $n$, $m$ being the integers. Obviously, the ADE model \cite{Cai:2007us} corresponds to the case $(n=0,m=0)$; the $\eta$HDE model \cite{Huang:2012xm} corresponds to $(n=0,m=-1)$, while the NADE model \cite{Wei:2007ty} corresponds to the same case but without considering a primordial part generated by inflation\cite{Huang:2012xm}; the CHDE model \cite{Huang:2012nz} corresponds to $(n=4, m=3)$.

In this note, we are going to investigate in detail a general class of holographic type dark energy models characterized by the IR cutoff given in Eq.(\ref{L0}). Our main purpose is to see the possible holographic type dark energy candidates for various cases of $(n, m)$ by requiring the corresponding models to be self-consistent and also consistent with the expansion history of the universe. In particular, we will show which cases are more favorable. In Sec.\,II, we first present a description on the general class of holographic type dark energy models, and then we make a detail investigation on various models and find out the possible candidates for the various choices $(n, m)$; In Sec.\,III, we will mainly focus on the single-parameter models similar to the $\Lambda$CDM model and perform the best-fit analysis by using recent cosmological observations; Some concluding remarks and discussions are given in Sec.\,IV.

\section{A General Class of holographic type dark energy Models }

Let us begin with the general characteristic length scale of the universe
 \begin{equation}
    L=\frac{1}{a^n(t)}\int_0^t dt'~a^m(t')=\frac{1}{a^n(t)}\int_0^a a'^m\frac{da'}{H'a'} ~,  \label{l}
 \end{equation}
where $H\equiv \dot{a}/a$ is the Hubble parameter and $\cdot$ denotes the derivative with respect to the cosmic time $t$. The corresponding holographic type dark energy
density is parameterized by
 \begin{equation}
    \rho_{de}=3d^2 M_p^2 L^{-2} ~, \label{rho}
 \end{equation}
where $d$ is a positive constant parameter. The fractional energy density is defined by
 \begin{equation}
    \Omega_{de}=\frac{\rho_{de}}{3M_p^2H^2}=\frac{d^2}{H^2L^2} ~. \label{frho}
 \end{equation}

For simplicity, let us consider a flat Friedmann-Robertson-Walker~(FRW) universe containing the holographic type dark energy and ambient constituent with constant
equation of state (EoS) $w_m$.  The Friedmann equation is given by
 \begin{equation}
    3M_p^2H^2=\rho_m +\rho_{de}  ~,
 \end{equation}
or in fractional energy densities
 \begin{equation}
    \Omega_{de}+\Omega_{m}=1  ~,   \label{fri}
 \end{equation}
where $ \Omega_{m}=\frac{\rho_{m}}{3M_p^2H^2}$. If there is no direct energy interchange, each energy component is conservative respectively, which results in
conservation equations given by
 \begin{equation}
    \dot{\rho}_{i}+3H(1+w_{i})\rho_{i}=0  ~ \label{ceq}
 \end{equation}
with $i=m,\, de$. By using Eqs.(\ref{l}), (\ref{rho}), (\ref{frho}) and (\ref{ceq}), the EoS of the holographic type dark energy is found to be
 \begin{equation}
    w_{de}=-1-\frac23n+\frac2{3d}\sqrt{\Omega_{de}}a^{m-n}  ~.   \label{wde}
 \end{equation}
Cosmic acceleration requires that $w_{de}<-\frac13$ recently, which indicates $n >-1$ or equivalently $n\geq 0$ as $n$ is taken to be integer under our present
consideration.

The conservation of the ambient constituent with constant $w_m$ leads to $\rho_m=C_1 a^{-3(1+w_m)}$, where $C_1$ is a constant coefficient. Combining with the
Friedmann equation and the definition of fractional energy densities, we have
 \begin{equation}
    \frac1{Ha}= \frac1{ \scriptstyle \sqrt{\frac{C_1}{3M_p^2}}} \sqrt{a^{(1+3w_m)}(1-\Omega_{de})}  ~.  \label{ha}
 \end{equation}
By using Eqs.(\ref{l}) and (\ref{frho}), we get
 \begin{equation}
    \int_0^a a'^m\frac{da'}{H'a'}=\frac{a^{n+1} d }{\sqrt{\Omega_{de}}Ha} ~.
 \end{equation}
Substituting Eq.(\ref{ha}) into above equation and taking derivative with respect to $a$ in both sides, we arrive at the differential equation of motion for
$\Omega_{de}$
 \begin{equation}
    \frac{d\Omega_{de}}{da}=\frac{\Omega_{de}}{a}(1-\Omega_{de})\left(3(1+w_m)+2n-\frac2d \sqrt{\Omega_{de}} a^{m-n} \right)  ~. \label{ode}
 \end{equation}

Obviously, under the transformation $a \to \frac{a}{a_0}, ~ d \to d a_0^{n-m}$, with taking $a_0$ as the present scale factor of the universe, the energy density
Eq.(\ref{rho}), the fractional energy density Eq.(\ref{frho}), the EoS Eq.(\ref{wde}) and the differential equation Eq.(\ref{ode}) are all invariant. Namely,
performing such transformation, all expressions keep the same, so we can set $a_0=1$. From now on, we adopt that the parameter $d$ has absorbed a factor $a_0^{n-m}$
and set $a_0=1$. Note that the subscript``0'' always indicates the present value of the corresponding quantity.

Considering that our universe has successively experienced the inflation during which $w_m\simeq-1$ (which indicates quasi-de Sitter expansion), the
radiation-dominated epoch during which $w_m= \frac13$, and the matter-dominated epoch during which $w_m= 0$ before turning to accelerated expansion recently, we are
actually  able to approximately study the behaviors of  $L$, thus the fractional density $\Omega_{de}$ in the early universe directly. Here, we will simply ignore
the effect due to the transition from one ambient-constituent-dominated epoch to another ambient-constituent-dominated epoch.

Let us define
 \begin{equation}
   \tilde{L}=\frac{1}{a^m(t)}\int_{0}^t dt'~a^m(t')=\left( \frac{a_i}{a} \right)^m \tilde{L}_i + \frac{1}{a^m(t)}\int_{t_{i}}^t dt'~a^m(t')~, \quad \tilde{L}_i=\frac{1}{a_i^m}\int_0^{t_i} dt'~a^m(t') \label{lw}
 \end{equation}
where the subscript $i$ denotes the beginning of some epoch under consideration. Obviously, we have  $L=a^{m-n}\tilde{L}$.

When the constituent with constant $w_m$ dominates the universe from $t_i$, we have approximately $H^2 \propto \rho_{m} \propto  a^{-3(1+w_m)}$ from Fridemann
equations, which results in the following consequences
 \begin{eqnarray}
    &&\frac{1}{a^m(t)}\int_{t_{i}}^t dt'~a^m(t')=\frac{1}{a^m(t)}\int_{a_{i}}^a a'^m\frac{da'}{H'a'} \nonumber \\
    &=&\left\{\large  \begin{array}{cc} \frac{1}{m+\frac{3(1+w_m)}{2}} \left( \frac1H-\frac1{H_{i} }\left(\frac{a_{i}}{a} \right)^m  \right)  & \quad  ~m > -\frac{3(1+w_m)}{2}  ~, \\
    \frac{1}{H} \ln \left(\frac{a}{a_{i}}\right) & \quad ~m = -\frac{3(1+w_m)}{2} ~, \\
    \frac{1}{\mid m+\frac{3(1+w_m)}{2}\mid} \left(\frac1{H_{i}}\left( \frac{a_{i}}{a} \right)^m- \frac1H \right)  & \quad ~ m < - \frac{3(1+w_m)}{2}  ~.\end{array}
       \right. \nonumber \\
    &\stackrel{a\gg a_{i}}{\simeq}&\left\{\large  \begin{array}{cc} \frac{1}{m+\frac{3(1+w_m)}{2}}  \frac1H     & \quad  ~m > -\frac{3(1+w_m)}{2}  ~, \\
    \frac{1}{H} \ln \left(\frac{a}{a_{i}}\right) & \quad ~m = -\frac{3(1+w_m)}{2} ~, \\
    \frac{1}{\mid m+\frac{3(1+w_m)}{2}\mid} \frac1{H_{i}}\left( \frac{a_{i}}{a} \right)^m  & \quad ~ m < - \frac{3(1+w_m)}{2}  ~.\end{array} \right.
    \label{lw2}
 \end{eqnarray}
Noticing that the approximation in the limit $a\gg a_{i}$ is due to the fact that $H^2 \propto  a^{-3(1+w_m)}$ and
 \begin{equation}
  \frac1{H_{i} }\left(\frac{a_{i}}{a} \right)^m ~~ {\Big /} ~~ \frac1H   = \left( \frac{a_{i}}{a} \right)^{m+\frac{3(1+w_m)}{2}} \label{lwneq}
 \end{equation}

When $m>0$, then $m > -\frac{3(1+w_m)}{2}$ all the time in the early universe. From Eqs.(\ref{lw}) and (\ref{lw2}), we have
 \begin{equation}
   \tilde{L}=\frac{1}{a^m(t)}\int_{0}^t dt'~a^m(t')=\left( \frac{a_i}{a} \right)^m \tilde{L}_i + \frac{1}{m+\frac{3(1+w_m)}{2}} \left( \frac1H-\frac1{H_{i}}\left(\frac{a_{i}}{a} \right)^m   \right)~
 \end{equation}
for the constituent with constant $w_m$ dominating the universe. During the inflation, the Hubble parameter $H$ is constant approximately and the universe expands
exponentially. Those terms with the factor $\left( \frac{a_i}{a} \right)^m$ will soon become negligible, we have $\tilde{L}\simeq\frac{1}{m+\frac{3(1+w_m)}{2}}
\frac1H$ approximately, which also results in  $\tilde{L}_e \sim O(1/H_e)$ at the end of the inflation. Approximately, for the radiation dominated epoch, the
initial value $\tilde{L}_i\sim \tilde{L}_e$ is at the order of $O(1/H_e)$. Referring to Eq.(\ref{lwneq}) for $m > -\frac{3(1+w_m)}{2}$ and reminding that the
expansion of the universe, we have $\tilde{L}\simeq\frac{1}{m+\frac{3(1+w_m)}{2}} \frac1H$ for radiation-dominated epoch during which $w_m=\frac13$ as well. Similar
result holds during the matter-dominated era.  By using $L=a^{m-n}\tilde{L}$ and Eq.(\ref{frho}), we get the approximate fraction of dark energy in early universe,
i.e.
 \begin{equation}
    \Omega_{de}\simeq \frac{d^2}4(2m+3w_m+3)^2a^{2(n-m)}~,   \label{frhom}
 \end{equation}
where $w_m\simeq-1$ during inflation, $w_m=\frac13$ in radiation-dominated epoch and $w_m=0$ in matter-dominated epoch respectively. For self-consistency, we need
$n > m$ to ensure that $\Omega_{de} \ll 1$ when $a \ll 1$, thus the ambient matter dominated the early universe. It is not difficult to prove that for $n > m$,
Eq.(\ref{frhom}) is the approximate solution of the differential equation of $\Omega_{de}$ under the limit $1-\Omega_{de} \simeq 1$ when $a \ll 1$ consistently.

For the case $m=0$, from Eqs.(\ref{lw}) and (\ref{lw2}), we simply have $\Omega_{de} \simeq d^2 a^{2n}\left( \ln \left(\frac{a}{a_{i}}\right)\right)^{-2}$ during
inflation. While the approximate solution in matter- or radiation-dominated epoch is also given by Eq.(\ref{frhom}). Thus $n > m$ is also required to be
self-consistent.

Let us now pay attention to the case $n=m\geq0$. In this case, we have $\Omega_{de}\equiv\frac{d^2}4(2m+3w_m+3)^2$ at the matter-dominated epoch, which is the exact
solution of the differential equation. Referring to the EoS of dark energy Eq.(\ref{wde}), we get $w_{de}\equiv w_m$ which means that the holographic type dark energy
tracks the dominated component and never dominates. This is of course unacceptable. Moreover, unless the parameter $d$ is small enough, the faction of dark energy
would be too large in early universe to be consistent with primordial nucleosynthesis (BBN) \cite{Olive:1999ij}. While by referring to the EoS Eq.(\ref{wde}), a
tiny $d$ would make the present EoS of dark energy $w_{de0}$ severely deviating from $-1$ and be inconsistent with recent observations \cite{Suzuki:2011hu}.

For the case $m<-2$, then $m < -\frac{3(1+w_m)}{2}$ holds all the time in the early universe. From Eqs.(\ref{lw}), (\ref{lw2}) and Eq.(\ref{lwneq}), we have in
early universe
 \begin{equation}
   \tilde{L}\simeq \left(\tilde{L}_i+\frac{1}{\mid m+\frac{3(1+w_m)}{2}\mid} \frac1{H_{i}}\right)\left( \frac{a_{i}}{a} \right)^m \gg \frac1H ~.
 \end{equation}
Combining with $L=a^{m-n}\tilde{L}$ and $H^2 \propto \rho_{m} \propto  a^{-3(1+w_m)}$, we get the fraction of dark energy in early universe when $a_i \ll a \ll1$
\begin{equation}
    \Omega_{de} = \frac{d^2}{H^2L^2}=( H^{-1}/ \tilde{L} )^2 d^2a^{2(n-m)}  \propto d^2(a_i)^{2|m|} a^{3(1+w_m)+2n}  ~. \label{frhom2}
 \end{equation}
As it is analyzed above that $n\geq0$ is required by recent cosmic acceleration, we then have $\Omega_{de} \ll 1$ when $a \ll 1$ consistently for $m < 0$ here. The scaling property of $\Omega_{de}$ with respect to $a$ can also be resulted from the differential equation of $\Omega_{de}$ under the limit $1-\Omega_{de}\simeq1$ when $a\ll 1$. This is because in light of the above equation the last factor in the differential equation Eq.(\ref{ode}) is found to be
 \begin{equation}
    \frac2d\sqrt{\Omega_{de}} a^{m-n}= \frac2{HL}a^{m-n} = 2H^{-1}/ \tilde{L} \ll 1 ~,  \label{tiny}
 \end{equation}
which is negligible small.

For the cases $m=-1$ and $m=-2$, we can also obtain consistently Eq.(\ref{frhom2}) for $n\geq 0$ during the inflation by following the same argument for the case $m<-2$. While for the radiation- and matter-dominated era, more attention is needed. Let us rewrite $\tilde{L}$ in the following way,
 \begin{equation}
      \tilde{L}=\frac{1}{a^m(t)}\int_{t_{b}}^{t_e} dt'~a^m(t')+ \frac{1}{a^m(t)}\int_{t_{e}}^t dt'~a^m(t')~, \label{tl12}
 \end{equation}
where subscripts $b$ and $e$ denote the beginning and the end of inflation respectively. From Eq.(\ref{lw2}), it is not difficult to find that the second term
$\frac{1}{a^m(t)}\int_{t_{e}}^t dt'~a^m(t')$ is approximately at order of $O(1/H)$ for the cases $m=-1$ and $m=-2$, while the first term is given by
 \begin{equation}
        \frac{1}{a^m(t)}\int_{t_{b}}^{t_e} dt'~a^m(t') \simeq \frac{1}{\mid m \mid} \frac{a}{H_{b}a_{b}}\left( \frac{a}{a_{b}} \right)^{\mid m \mid -1} ~.
 \end{equation}
In order to solve the horizon problem \cite{Guth:1980zm}, the inflation is required to last enough time to make
 \begin{equation}
    \frac1{H_b a_b} > \frac1{H_0 a_0}  ~.\label{neq}
 \end{equation}
Moreover, due to the fact that $\frac1{Ha}$ has been growing as $a^{(1+3w)/2}$ until recent cosmic acceleration (from then on, $w<-\frac13$), we generally have
$\frac1{H_0 a_0} \gg \frac1{H a}$ when $a_e\ll a \ll a_0$. Thus for $m\leq-1$, we yield
 \begin{equation}
       \frac{1}{\mid m \mid} \frac1{H_{b}a_{b}}\left( \frac{a}{a_{b}} \right)^{\mid m \mid -1} > \frac1{H_0 a_0} \gg \frac1{H a}~,  \label{neq2}
 \end{equation}
which indicates that for the cases $m=-1$ and $m=-2$, the first term in Eq.(\ref{tl12}) is much larger than the second term. Thus we approximately have
 \begin{equation}
      \tilde{L}\simeq \frac{1}{a^m(t)}\int_{t_{b}}^{t_e} dt'~a^m(t') \gg \frac1{H } ~
 \end{equation}
in the radiation- and matter-dominated epoch. Therefore, Eq.(\ref{frhom2}) also holds at these two epoches for $n\geq0$ and  $m=-1$ or $m=-2$.

It is noticed that for the case $n=0$ and $m<0$ the characteristic scale $L$ will be dominated by the primordial part generated in inflation with referring to
Eqs.(\ref{l}), (\ref{lw2}), (\ref{neq2}), i.e.
 \begin{equation}
        L=\int_{0}^{t} dt'~a^m(t')\simeq L_{\rm prim}=\int_{t_{b}}^{t_e} dt'~a^m(t') \simeq \frac{1}{\mid m \mid} \frac{1}{H_{b}a_{b}}\left( \frac{1}{a_{b}} \right)^{\mid m \mid -1}
 \end{equation}
in the radiation- and matter-dominated epoch. Then the dark energy behaves almost like a cosmological constant during these two epoches with
 \begin{equation}
    \rho_{de}\simeq 3d^2 M_p^2 L_{\rm prim}^{-2}~.
 \end{equation}
Correspondingly,  the fractional energy density of dark energy scales as
 \begin{equation}
    \Omega_{de}\simeq \frac{d^2}{H^2L_{\rm prim}^2} \propto \frac{d^2}{L_{\rm prim}^2}a^{3(1+w_m)} ~.
 \end{equation}
The $\eta$HDE model corresponding to the case $(n=0,m=-1)$ has been investigated in \cite{Huang:2012xm}. In the cases with $m<0$, the observation constraints on the parameter $d$ is very weak and the models seemly have only one effective parameter, i.e. the ratio $L/d\sim O(H_0^{-1})$. In the $\eta$HDE model, the parameter $d$ can take value in a normal order \cite{Huang:2012xm}. While for the case $m<-1$, the parameter $d$ seems to take very large value due to the very large factor in $L$, i.e. $\left( \frac{1}{a_{b}}\right)^{\mid m \mid -1}$, in order to make the ratio $L/d$ in the right order.

In summary, $n \geq 0$ is required by the recent cosmic accelerated expansion. For $n \geq 0$ and $m<0$, the fraction of dark energy density scales as $\Omega_{de}
\propto  a^{3(1+w_m)+2n}$ with a tiny proportionality coefficient in the early universe when $a \ll 1$. Since the proportionality coefficient can not be determined
only by the parameter $d$,  there are in general two model parameters. They may be chosen to be the parameter $d$ and the present fraction of dark energy $\Omega_{de0}$. However,
if $n=0$, the holographic type dark energy would behave almost like a cosmological constant and there seems to be only one effective parameter $\Omega_{de0}$. Since the
total energy density $\rho=3M_p^2H^2 \propto a^{-3(1+w_m)}$ approximately when the constituent with constant $w_m$ dominates the universe, we have $\rho_{de}\propto
a^{2n}$. Thus $w_{de}\simeq -1-\frac{2n}{3}$ in the early universe.

For $n > m > 0$, we have $\Omega_{de}\simeq \frac{d^2}4(2m+3w_m+3)^2a^{2(n-m)}$ in the early universe when $a \ll 1$.  For $n>m=0$, we also have
$\Omega_{de}\simeq\frac{d^2}4(2m+3w_m+3)^2a^{2(n-m)}$ in the matter- or radiation-dominated epoch, while $\Omega_{de} \simeq d^2 a^{2n}\left(
\ln\left(\frac{a}{a_{i}}\right)\right)^{-2}$ during inflation.

It is seen that for all cases $n > m \geq 0$ the fraction of dark energy can be ignored naturally in early universe when $a \ll 1$ as long as the parameter $d$
takes a normal order value.  Moreover, due to such analytic feature, we can use the approximate solution at some $a_{\rm ini}\ll 1$ in matter-dominated epoch as the
initial condition to solve the equation of motion for $\Omega_{de}$. Noticing that once $d$ is given, the present fractional energy density $\Omega_{de}(a=1)$ can
be obtained. So the model is a single-parameter model like the $\Lambda$CDM model. Substituting this approximate solution to the EoS of dark energy Eq.(\ref{wde}),
we get
\begin{equation}
w_{de}=-\frac23(n-m)+w_m
\end{equation}
in the ambient-constituent-dominated epoch. Referring to Eq.(\ref{wde}), $w_{de}$ will transit to $-1-\frac{2n}3$ due to the expansion of the universe. Obviously,
for models with $n=m+1 >0$, the EoS of the dark energy in the inflation, radiation- and matter-dominated epoch are $-\frac53$, $-\frac13$ and $-\frac23$
respectively, and transit to $-1-\frac{2n}3$ eventually. Therefore, the EoS of dark energy in such kinds of models transit from $w_{de}<-1$ to $w_{de}>-1$ and back
to $w_{de}<-1$ during the universe expansion.

The choices $n=m\geq0$ are abandoned because the dark energy cannot dominate the universe forever and there might be too large fraction of dark energy in early universe. While the choices $m> n \geq 0$ are forbidden by the self-consistent requirement that $\Omega_{de}\ll 1 $ when $a\ll 1$.

We are going to mainly focus on the single-parameter models corresponding to the cases $n>m \geq 0$. In next section, we will perform the best-fit analyses on such kinds of models by using recent observations. As the $\eta$HDE model \cite{Huang:2012xm} corresponding to the case $(n=0, m=-1)$ has only one effective parameter, we shall also take it into consideration for a comparison.

\section{Observational constraints on single-parameter models}

In this section,  we are going to mainly focus on the single-parameter models corresponding to the cases $n>m \geq 0$ and perform best-fit analyses on them by using
recent cosmological observations including the Union2.1 compilation of 580 supernova Ia (SNIa) data \cite{Suzuki:2011hu}, the parameter $A$ from BAO
measurements\cite{Eisenstein:2005su} and the shift parameter $R$ from CMB measurements \cite{Komatsu:2010fb}. The observational data and analysis method are given
in Appendix A.

As the cosmological observations mainly come from the epoch when the fraction of radiation energy is tiny, we then simply consider a flat
Friedmann-Robertson-Walker~(FRW) universe containing only dark energy and matter with $w_m=0$. Using $a=\frac1{1+z}$ with $z$ the redshift, we can rewrite
Eq.(\ref{ode}) as
 \begin{equation}
 \frac{d\Omega_{de}}{dz}=-\frac{\Omega_{de}(1-\Omega_{de})}{1+z}\left (2n+3-\frac2d\sqrt{\Omega_{de}}(1+z)^{n-m}\right)~ \label{odez}
 \end{equation}
and the approximate solution for the cases $n>m \geq 0$ in the matter-dominated epoch is given by
 \begin{equation}
    \Omega_{de} \simeq \frac{d^2}4(2m+3)^2(1+z)^{-2(n-m)}~,  \label{frhoz}
 \end{equation}
which can be chosen as an approximate solution at some $z_{ini}$. Namely, we may take $\Omega_{de}(z_{ini})=\frac{d^2}4(2m+3)^2 (1+z_{ini})^{-2(n-m)} $ as the
initial condition to solve the differential equation Eq.(\ref{odez}). The solution depends weakly on the choice of $z_{ini}$ in a wide range, since $\Omega_{de}$ is
tiny and scales as $(1+z)^{-2(n-m)}$ when $z\gg 1$. Here, we simply take $z_{ini}=2000$ at which matter dominates the universe well.

As the $\eta$HDE model \cite{Huang:2012xm} corresponding to the case $(n=0, m=-1)$ behaves as a single effective parameter model, we would like to take it into
consideration for a comparison. Reminding that there are actually two parameters in this model, we may choose $d$ and $\Omega_{m}(z=0)$ as two parameters. Taking
$\Omega_{de}(z=0)=1-\Omega_{m}(z=0)$ as the boundary condition, we are able to solve the differential equation Eq.(\ref{odez}).

From the Friedmann equation and conservative equations, we have
 \begin{equation}
 H^2(z)=H_0^2\Omega_{m0}(1+z)^3 +H^2(z)\Omega_{de}(z)~.
 \end{equation}
Equivalently,
 \begin{equation}
 E(z)\equiv {H(z)\over H_0}=\left(\Omega_{m0}(1+z)^3\over 1-\Omega_{de}(z)\right)^{1/2}  ~. \label{Ez}
 \end{equation}
Substituting the results of $\Omega_{m0}=1-\Omega_{de}(z=0)$ and $\Omega_{de}(z)$ by solving Eq.(\ref{odez}) into Eq .(\ref{Ez}), the function $E(z)$ can be
obtained.

Of course, it is impossible to perform the best-fit analyses on all single-parameter models corresponding to the cases $n>m \geq 0$. However, we can still learn something from the best-fit analyses on a sample of models. Here, we are going to focus on the models with subjecting to $ 0\leq n \leq 7$ as well as $ max(0, n-4)\leq m < n$ and to figure out some general results. The $\eta$HDE model corresponding to the case $(n=0, m=-1)$ is also included.

In Table [\ref{fit1}], we present the best-fit $\chi^2$ results  by using only the Union2.1 compilation of 580 supernova Ia (SNIa) data \cite{Suzuki:2011hu}. For
comparison, we also fit the $\Lambda$CDM model to the same observational data, and find that the minimal $\chi^2_{\Lambda {\rm CDM}}=562.227$ for the best fit
parameter $\Omega_{m0}=0.278$. Obviously, the $\Lambda$CDM model fits to the SNIa data best. It is interesting to note that the best-fit result of the $\eta$HDE
model is the same as the $\Lambda$CDM model. Actually, the $\eta$HDE model reduces to the $\Lambda$CDM model when model parameter $d\to \infty$ \cite{Huang:2012xm}.
From Table [\ref{fit1}], when focusing on the single-parameter models with $n>m \geq 0$, we see that models with $n-m=1$ have much smaller best-fit $\chi^2$
functions than models with $n-m>1$ for the given $n$. Also for the given $n-m$, the best-fit $\chi^2$ function increases with $n$ .

\begin{table}
\caption{ The minimum of $\chi^2$  by using only the Union2.1 SNIa data; for comparison, $\chi^2_{\Lambda {\rm CDM}}=562.227$. }
\begin{center}
\label{fit1}
\begin{tabular}{|c|c|c|c|c|c|c|c|c|}
\hline                 &     n=0         &     n=1        &   n=2          &   n=3          &   n=4          &   n=5          &   n=6          &   n=7            \\
\hline      m=n-1      &     562.227     &   562.242      &   562.657      &   563.212      &   563.751      &   564.244      &   564.686      &   565.083          \\
\hline      m=n-2      &     -           &      -         &   568.794      &   570.957      &   572.680      &   574.088      &   575.264      &   576.265        \\
\hline      m=n-3      &     -           &      -         &      -         &   583.096      &   585.780      &   587.928      &   589.696      &   591.183       \\
\hline      m=n-4      &     -           &      -         &      -         &      -         &   600.560      &   603.214      &   605.394      &   607.223      \\
\hline
\end{tabular}
\end{center}
\end{table}
\begin{table}
\caption{ The best-fit results of some models with $n-m=1$ by using only the Union2.1 SNIa data }
\begin{center}
\label{fit2}
\begin{tabular}{|c|c|c|c|c|c|c|c|c|}
\hline                 &     n=0         &     n=1        &   n=2          &   n=3          &   n=4          &   n=5          &   n=6          &   n=7          \\
\hline      $\chi^2$   &     562.227     &   562.242      &   562.657      &   563.212      &   563.751      &   564.244      &   564.686      &   565.083       \\
\hline        $d$      &   $ \gtrsim O(10)$ &    0.874       &   0.460        &   0.307        &   0.229        &   0.181        &   0.150        &   0.127        \\
\hline  $\Omega_{m0}$  &     0.278       &    0.246       &   0.272        &   0.287        &   0.298        &   0.305        &   0.311        &   0.316        \\
\hline     $w_{de0}$   &    $\sim$ -1.000      &    -0.997      &   -1.098       &   -1.167       &   -1.223       &   -1.266       &  -1.302        &   -1.332      \\
\hline
\end{tabular}
\end{center}
\end{table}
\begin{table}
\caption{ The best-fit $\chi^2$ by using SNIa+BAO+CMB data sets; for comparison, $\chi^2_{\Lambda {\rm CDM}}=562.531$.  }
\begin{center}
\label{fit3}
\begin{tabular}{|c|c|c|c|c|c|c|c|c|}
\hline                 &     n=0         &     n=1        &   n=2          &   n=3          &   n=4          &   n=5          &   n=6          &   n=7            \\
\hline      m=n-1      &     562.383    &   566.363      &   562.872      &   563.738      &   565.723      &   567.922      &   570.057      &   572.039         \\
\hline      m=n-2      &     -           &      -         &   636.152      &   650.851      &   661.528      &   669.697      &   676.188      &   681.495        \\
\hline      m=n-3      &     -           &      -         &      -         &   737.659      &   749.054      &   757.655      &   764.430      &   769.934       \\
\hline      m=n-4      &     -           &      -         &      -         &      -         &   813.612      &   821.699      &   828.098      &   833.314      \\
\hline
\end{tabular}
\end{center}
\end{table}
\begin{table}
\caption{ The best-fit results at 1 $\sigma$ (68.3\%) and 2 $\sigma$ (95.4\%) confidence levels with SNIa+BAO+CMB data sets; for $\Lambda$CDM  model,
$\chi^2_{\Lambda {\rm CDM}}=562.531$ and $\Omega_{m0}=0.273^{+0.014}_{-0.013}~ ^{+0.028}_{-0.026} $.}
\begin{center}
\label{fit4}
\begin{tabular}{|c|c|c|c|c|c|c|}
\hline                 &   $(n=0, m=-1)$                                          &         $(n=1, m=0)$                                &         $(n=2, m=1)$                                &         $(n=3, m=2)$                               &        $(n=4, m=3)$                                 \\  
\hline     $\chi^2$    &   562.383                                                &    566.363                                          &   562.872                                           &   563.738                                          &    565.723                                          \\  
\hline      $d$        &   $34.5^{+\infty}_{-27.1}~ ^{+\infty}_{-29.4} $          &    $0.833^{+0.018}_{-0.018}~ ^{+0.037}_{-0.036}$    &    $0.459^{+0.009}_{-0.009}~ ^{+0.017}_{-0.017}$    &   $0.310^{+0.005}_{-0.005}~ ^{+0.010}_{-0.011}$    &   $0.232^{+0.004}_{-0.004}~ ^{+0.007}_{-0.008} $    \\  
\hline  $\Omega_{m0}$  &   $0.272^{+0.022}_{-0.021}~ ^{+0.036}_{-0.033} $         &    $0.267^{+0.013}_{-0.013}~ ^{+0.026}_{-0.024}$    &    $0.275^{+0.013}_{-0.012}~ ^{+0.025}_{-0.024}$    &   $0.282^{+0.012}_{-0.012}~ ^{+0.025}_{-0.023}$    &    $0.288^{+0.013}_{-0.012}~ ^{+0.025}_{-0.023} $   \\  
\hline    $w_{de0}$    &   $-0.984^{+0.060}_{-0.016}~ ^{+0.096}_{-0.016} $        &    $-0.981^{+0.009}_{-0.009}~ ^{+0.018}_{-0.018}$   &    $-1.095^{+0.013}_{-0.013}~ ^{+0.026}_{-0.025}$   &   $-1.176^{+0.015}_{-0.015}~ ^{+0.031}_{-0.031}$   &    $-1.237^{+0.018}_{-0.018}~ ^{+0.035}_{-0.035} $  \\  
\hline
\end{tabular}
\end{center}
\end{table}

\begin{figure}
\centerline{
\includegraphics[width=4.2cm,height=7.5cm]{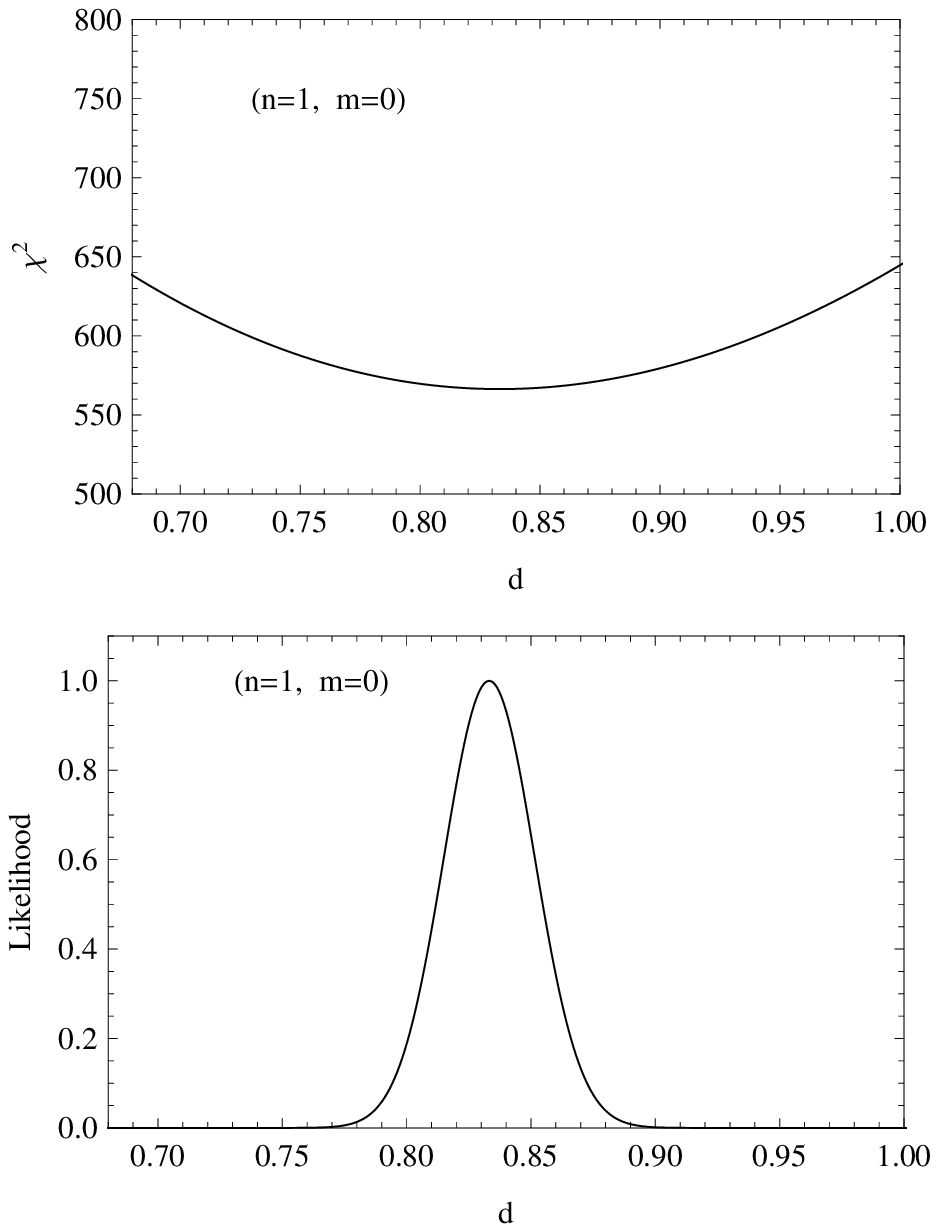}
\includegraphics[width=4.2cm,height=7.5cm]{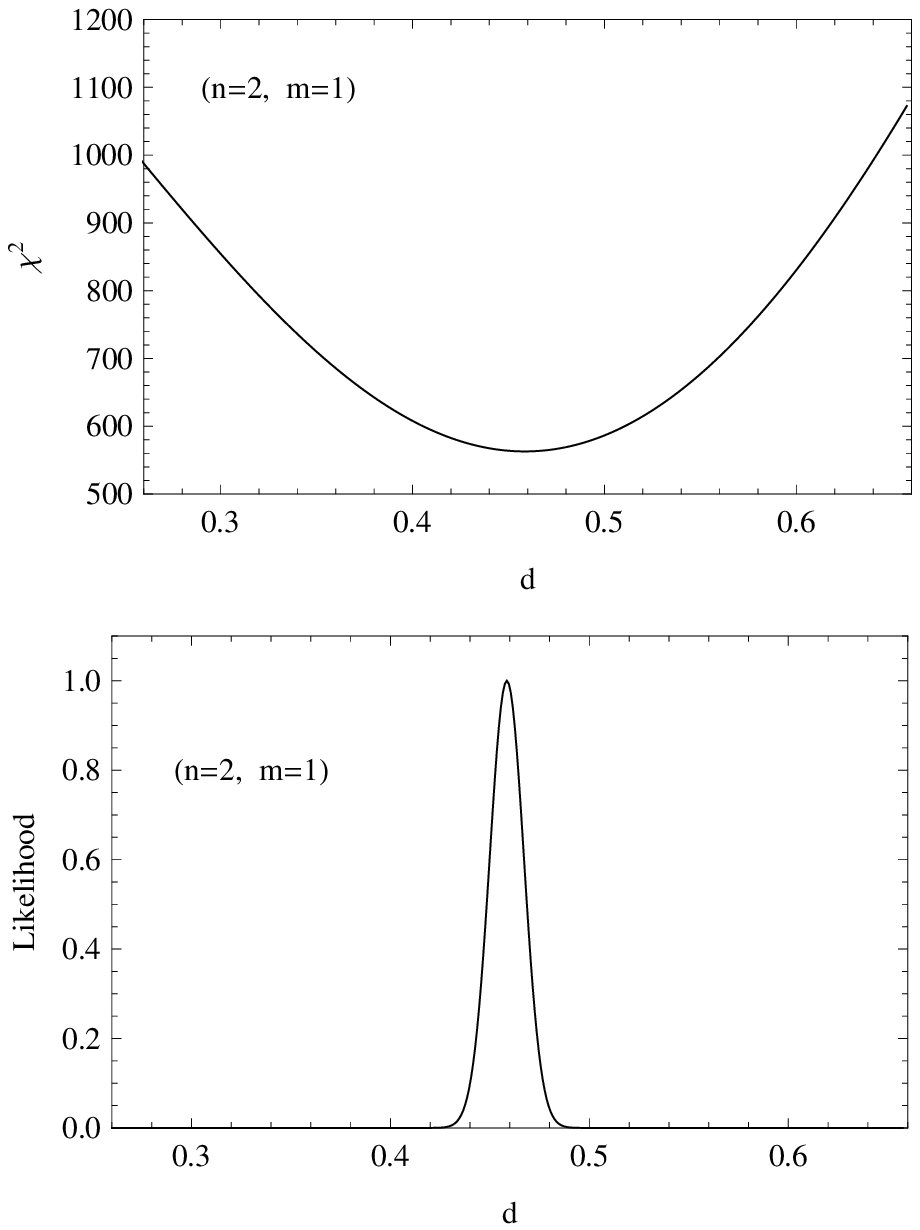}
\includegraphics[width=4.2cm,height=7.5cm]{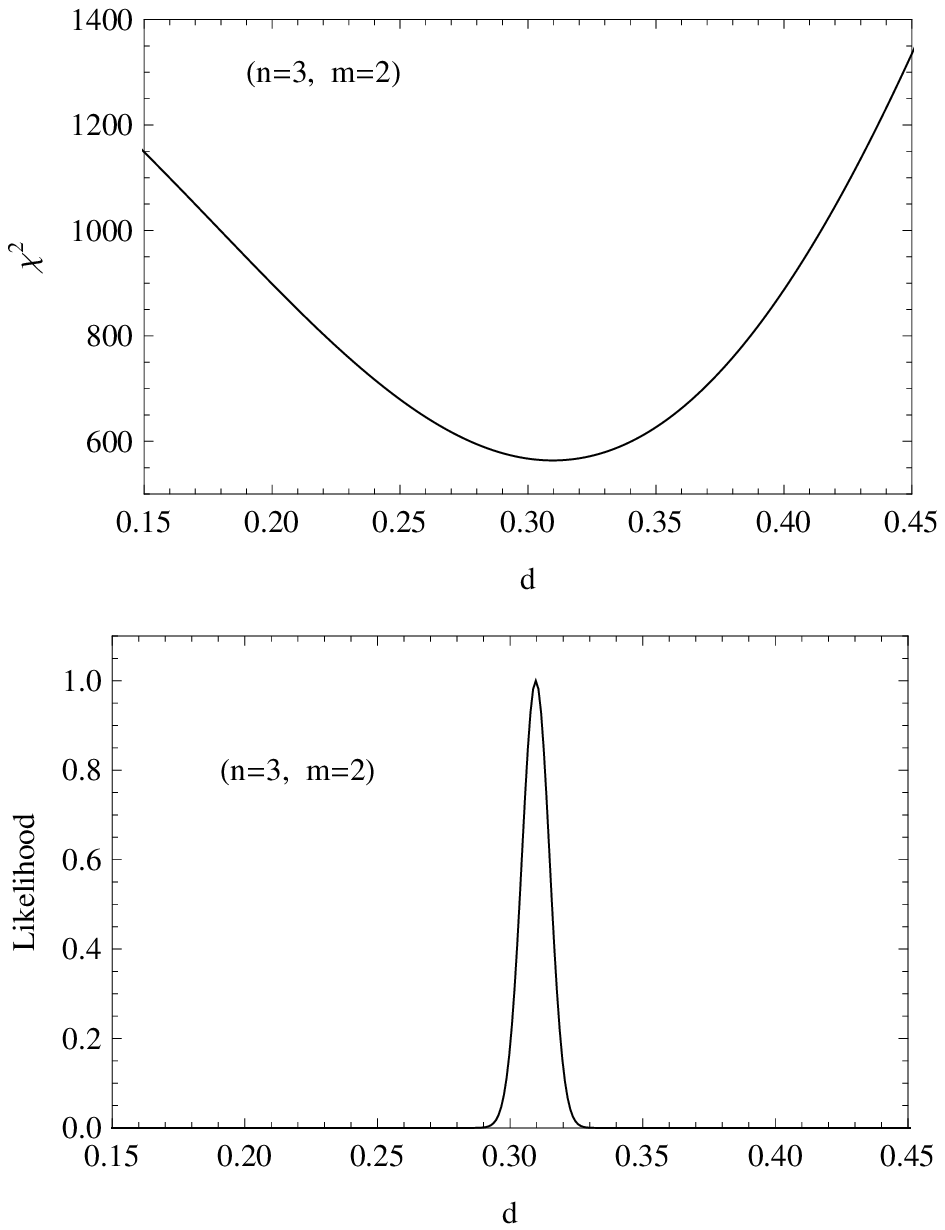}
\includegraphics[width=4.2cm,height=7.5cm]{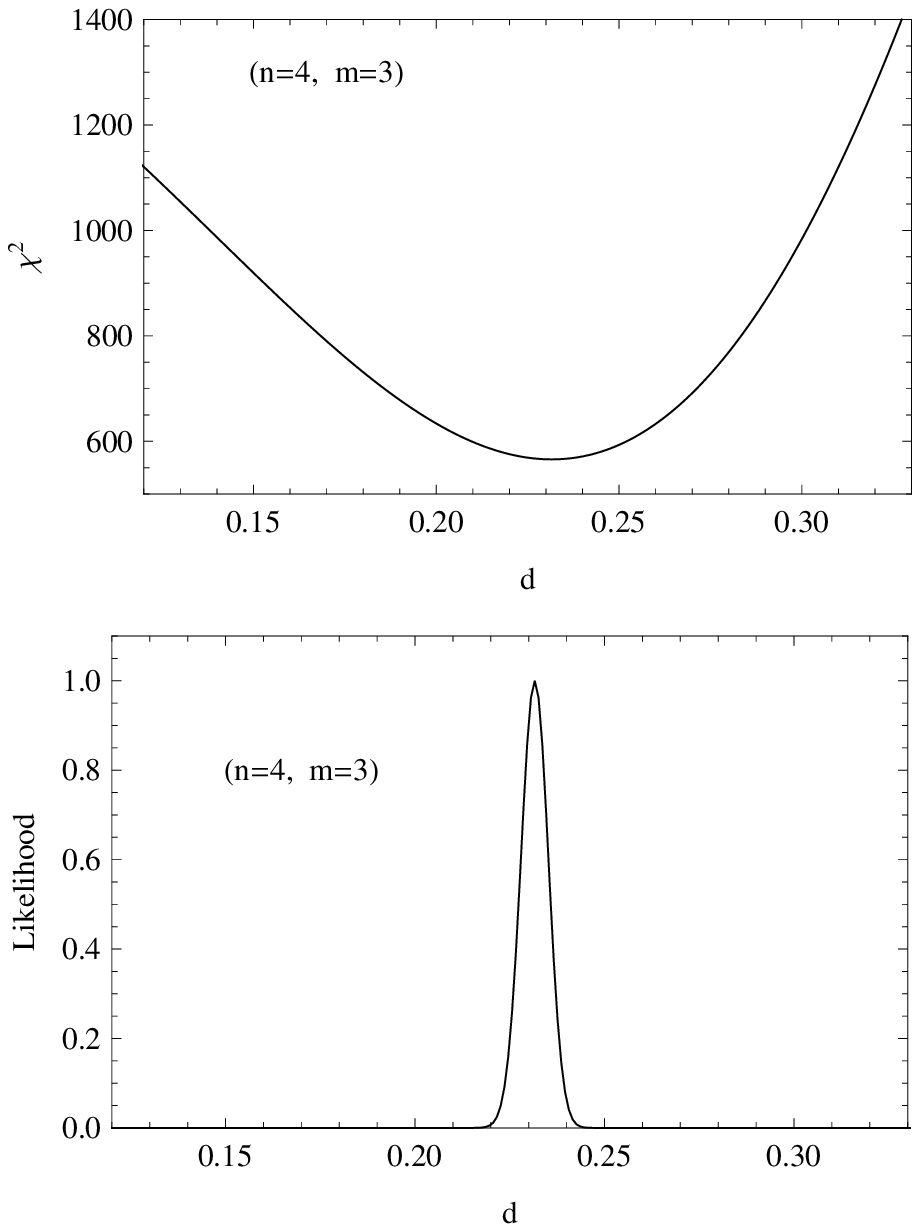}
} \caption{\label{fig1} The $\chi^{2}$ functions and corresponding likelihoods of models by using SNIa+BAO+CMB data sets.}
\end{figure}

\begin{figure}
\centerline{\includegraphics[width=13cm, height=5cm]{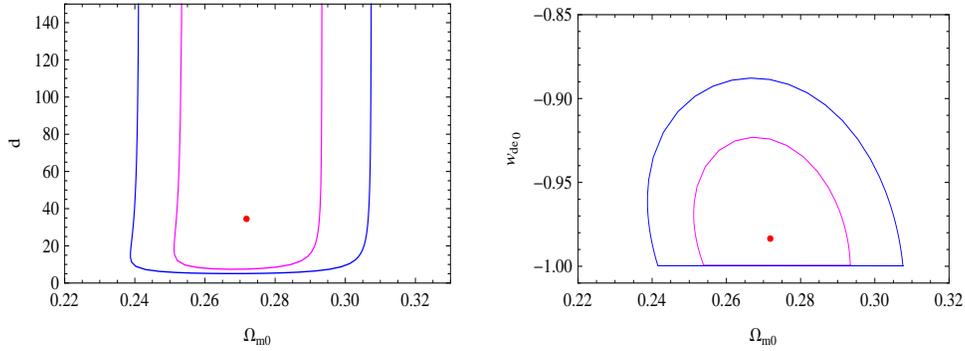}} \caption{\label{fig2}  Probability contours at 68.3\% and 95.4\% confidence levels of $\eta$HDE
model with $(n=0, m=-1)$ ; the best-fit values are $d=34.5$ and $\Omega_{m0}=0.272$ by using SNIa+BAO+CMB data sets.}
\end{figure}

It is noticed that the single-parameter models characterized by the conformal-age-like length scale with $n-m=1$ seem to be more favored, we shall pay more attention to those models. The best-fit results for those
models with some corresponding quantities are summarized in Table [\ref{fit2}]. We can also see that the present EoS of dark energy $w_{de0}$ deviating from $-1$
become much more for larger $n$. Among the single-parameter models $(n\geq1)$ the present fraction of matter $\Omega_{m0}$ is increased with larger $n$. It is
interesting to note that the parameter $d$ is compatible with $1/n$ in Table [\ref{fit2}]. This can be enlightened by the EoS given in Eq.(\ref{wde}). The present
EoS of dark energy is $w_{de0}(a=1)=-1-\frac23n+\frac2{3d}\sqrt{\Omega_{de0}}$. As $\Omega_{de0}\lesssim 1$, it leads to $d$ being compatible with $1/n$ in order to
get $w_{de0}$ deviating less from $-1$ which seems to be more favored by the observations.

The complimentary data sets from different cosmological observations often constrain dark energy models better. Therefore, we would like to perform joint analyses
on the same sample of models by using the Union2.1 compilation of 580 supernova Ia (SNIa) data \cite{Suzuki:2011hu}, the parameter $A$ from BAO measurements
\cite{Eisenstein:2005su} and the shift parameter $R$ from CMB measurements \cite{Komatsu:2010fb}. In Table [\ref{fit3}], we show the best-fit $\chi^2$ results for
those models. For the given $n$, we again conclude that models with $n-m=1$ are more favored.  For the given $n-m$, the best-fit $\chi^2$ function increases
generally with $n$ except that the best-fit $\chi^2$ of the model with $(n=1,m=0)$ is larger than the model with $(n=2,m=1)$. This is mainly because the model with
$(n=1,m=0)$ favors much smaller fraction of matter and goes more against the BAO and CMB observations which are more sensitive to the present fraction of matter
than the SNIa data.

The best-fit analysis indicates that the models characterized by the conformal-age-like length $L=\frac1{a^n(t)}\int_0^t dt'~a^{n-1}(t')$ with small $n$ are more favored from the cosmological observations. In Table \ref{fit4}, we present the best-fit results of models with $n-m=1$ as well as $0\leq n\leq4$ at the
$68.3\%~(95.4\%)$ confidence level by using SNIa +BAO+CMB data sets. For the single-parameter model, $\Delta\chi^2\equiv\chi^2-\chi_{min}^2\leq 1.0 ~(4.0)$ is used
to constrain the model parameter at the $68.3\%~(95.4\%)$ confidence level, while $\Delta\chi^2\leq 2.3 ~(6.17)$ is used for the two-parameter model. The likelihoods or probability contours of these models are plotted in figure \ref{fig1} and \ref{fig2}. From Table
\ref{fit4}, we find that the present EoS of the dark energy $w_{de0}$ is around $-1$ for model with $(n=1,m=0)$, while for models with $n> 1$ the present EoS of
dark energy are significantly below $-1$ at two $2\sigma$ level. It is also noticed that the EoS of dark energy in the $\eta$HDE model has $w_{de0} \agt -1$ and
slightly deviates from $-1$. In Table \ref{fit4}, just as in Table \ref{fit2}, one notices again that the model parameter $d$ is around $d\sim O(1/n)$ in those four
single-parameter models. It seems to hold for models with $n=m+1>0$.  By referring to Eq.(\ref{frhom}), the holographic type dark energy in such single-parameter models
characterized by the conformal-age-like length $L=\frac1{a^n(t)}\int_0^t dt'~a^{n-1}(t')$ with small positive $n$ can be ignored in early universe as
$\Omega_{de}\simeq \frac{d^2}4(2n+3w_m+1)^2a^2$ when $a\ll1$, and the conformal-age-like length is given by
$L\propto \frac1{aH}$.

\section{conclusion and discussion}

We have investigated a general class of holographic type dark energy models described by the characteristic length scale $L=\frac1{a^n(t)}\int_0^t dt'~a^m(t')$ with
integers $n$, $m$.  It has been shown that the recent cosmic accelerated expansion requires $n \geq 0$. For $n \geq 0$ and $m<0$, we have the fraction of dark
energy density scales as $\Omega_{de} \propto  a^{3(1+w_m)+2n}$ with a tiny proportionality coefficient and  $w_{de}\simeq -1-\frac{2n}{3}$ in the early universe
when $a \ll 1$. There are in general two model parameters which can be chosen as the parameter $d$ and the present fraction of dark energy $\Omega_{de0}$. For $n=0$
and $m<0$, the holographic type dark energy has been found to behave like a cosmological constant and there is actually only one effective parameter $\Omega_{de0}$. For
$n > m \geq  0$, we have $\Omega_{de}\simeq \frac{d^2}4(2m+3w_m+3)^2a^{2(n-m)}$ and $w_{de}\simeq-\frac23(n-m)+w_m$ in the early universe when $a \ll 1$ except that
$\Omega_{de} \simeq d^2 a^{2n}\left( \ln\left(\frac{a}{a_{i}}\right)\right)^{-2}$ during inflation for $m=0$. For those cases, the fraction of dark energy can be
ignored in early universe when $a \ll 1$ as long as the parameter $d$ takes a normal value. Moreover, due to the analytic feature in early universe, the models with
$n> m \geq  0$ have been found to be single-parameter models like the $\Lambda$CDM model. Particularly, the EoS of the dark energy in the models with $n=m+1>0$ have
been shown to transit from $w_{de}>-1$ in the radiation- and matter-dominated epoch to $w_{de}<-1$ eventually. It has been demonstrated that the choices $n=m\geq0$
should be abandoned as the dark energy cannot dominate the universe forever and there might be too large fraction of dark energy in early universe, while the
choices $m> n \geq 0$ must be forbidden from the self-consistent requirement that $\Omega_{de}\ll1 $ when $a\ll 1$.

It is worth to point out that although the energy density of the dark energy in those self-consistent models grows or falls more slowly than the energy density of
radiation and matter, the dark energy  cannot dominate the early universe due to the much smaller initial density, e.g. for the models with $n=m+1>0$, we have
$\rho_{de}(a_e)=\Omega_{de}(a_e)3M_p^2H_e^2 \sim \rho_{r}(a_e) \cdot a_e^2 \ll \rho_{r}(a_e) $ at the end of the inflation at $a_e\ll1$. But the dark energy will
eventually dominate the universe and be responsible for the recent cosmic acceleration.  Interestingly, for models with $n > m \geq  0$, the pre-inflation part of
the characteristic length $L$ is redshifted by the inflation which results in that $\Omega_{de}(a_e)$ is determined by the model parameter $d$ and the inflation
approximately. This means that the coincident problem of dark energy might be solved by the inflation naturally in these models.

The model with $(n= 0, m=-1)$ and the single-parameter models corresponding to cases $n >m \geq 0$ have been studied by using the recent cosmological observations. It has been shown that the five models with $(n= 0, m=-1)$, $(n= 1, m=0)$, $(n= 2, m=1)$, $(n= 3,
m=2)$ and $(n=4, m=3)$ fit to observations well. In the case $(n= 0, m=-1)$, the characteristic length scale $L$ is dominated by the primordial part generated by the inflation, which resulted in small and almost constant dark energy density \cite{Huang:2012xm}. While for the single-parameter models characterized by the conformal-age-like length $L=\frac1{a^n(t)}\int_0^t dt'~a^{n-1}(t')$ with $n>0$, it can be seen that the characteristic length scale behaves as $L\propto\frac1{Ha}$ in early universe, and the best-fit analysis leads the model parameter $d$ to be $d\sim O(1/n)$. Thus, the holographic type dark energy in such single-parameter models characterized by the conformal-age-like length $L=\frac1{a^n(t)}\int_0^t dt'~a^{n-1}(t')$ with small positive $n$ can be ignored in early universe as $\Omega_{de}\simeq \frac{d^2}4(2n+3w_m+1)^2a^2$ when $a\ll1$. The best-fit analysis has shown that the models characterized by the conformal-age-like length $L=\frac1{a^n(t)}\int_0^t dt'~a^{n-1}(t')$ with small $n $  are
more favored from the cosmological observations.

\section*{Acknowledgements}
We would like to thank R.G. Cai, M. Li, Y. Ling, J.X. Lu for useful discussions. The author (Z.P.H) would like to thank M. Q. Huang and M. Zhong for their helpful
support. This work is supported in part by the National Basic Research Program of
China (973 Program) under Grants No. 2010CB833000; the National Nature Science Foundation of China (NSFC) under Grants No. 10975170, 10975184, 10947016. \\

\appendix

\section{ Observational data and analysis method}

In this appendix, we present the method used for the best-fit analysis on the observational data including Type Ia Supernovae (SNIa), Baryon Acoustic Oscillations
(BAO), Cosmic Microwave Background (CMB) radiation.

For only Type Ia Supernovae (SNIa) observation, the likelihood function is given by
 \begin{eqnarray}
  \cal{L }_{\rm SN} &=& \exp[-\tilde{\chi}^2_{\rm SN}/2]~.
\end{eqnarray}

For the three independent observations, the likelihood function of a joint analysis is
 \begin{eqnarray}
  \cal{L }&=& \cal{L }_{\rm SN} \times \cal{L }_{\rm BAO}  \times \cal{L }_{\rm CMB}  \nonumber \\
    &=& \exp[-(\tilde{\chi}^2_{\rm SN}+\chi_{\rm BAO}^2+\chi_{\rm CMB}^2)/2]~.
\end{eqnarray}

The model parameters yielding a minimal $\sum_{i}\chi_i^{2}$ and a maximal $\cal{L }$ will be favored by the observations. In the following, we present the
calculation for the various  $\chi_i^{2}$ of each observational data set.

\subsection{Type Ia Supernovae (SN Ia)}

We consider the Supernova Cosmology Project (SCP) Union2.1 compilation \cite{Suzuki:2011hu}, which compiles the distance modulus $\mu_{\rm obs}(z_i)$ of 580
supernovae. The theoretical distance modulus is defined as
\begin{equation}
\mu_{\rm th}(z_i)\equiv 5 \log_{10} {D_L(z_i)} +\mu_0   ~,
\end{equation}
in which the parameter $\mu_{0}$ is a nuisance parameter but is independent of the data. The Hubble-free luminosity distance is given by
\begin{equation}
 D_{L}(z)=\left(1+z\right)\int_{0}^{z}\frac{dz'}{E(z')} ~,
\end{equation}
with $E(z) \equiv H(z)/H_{0}$.

The $\chi^2$ for the SNIa data is given by
\begin{equation}
\chi^2_{\rm SN}=\sum\limits_{i}{[\mu_{\rm obs}(z_i)-\mu_{\rm th}(z_i)]^2\over \sigma_i^2} ~, \label{ochisn}
\end{equation}
where $\mu_{\rm obs}(z_i)$ and $\sigma_i$ are the observed value and the corresponding 1$\sigma$ error of distance modulus for each supernova, respectively. To
reduce the effect of nuisance parameter $\mu_{0}$ \cite{Perivolaropoulos:2004yr}, one can expand $\chi^2_{\rm SN}$ with respect to $\mu_0$ as follows
\begin{equation}
\chi^2_{\rm SN}=A-2\mu_0 B+\mu_0^2 C ~, \label{ochisn2}
\end{equation}
where
\begin{equation}
A=\sum\limits_{i}{[\mu_{\rm obs}(z_i)-\mu_{\rm th}(z_i;\mu_{0}=0)]^2\over \sigma_i^2}  ~,
\end{equation}
\begin{equation}
B=\sum\limits_{i}{\mu_{\rm obs}(z_i)-\mu_{\rm th}(z_i;\mu_{0}=0)\over \sigma_i^2} ~,
\end{equation}
\begin{equation}
C=\sum\limits_{i}{1\over \sigma_i^2} ~.
\end{equation}
Evidently, $\chi_{\mathrm{SN}}^{2}$ has a minimum as
\begin{equation}
\tilde{\chi}^2_{\rm SN}=A-\frac{B^2}C ~, \label{tchi2sn}
\end{equation}
which is independent of $\mu_0$. Since $\tilde{\chi}^2_{\rm SN, min}=\chi^2_{\rm SN, min}$, we will adopt $\tilde{\chi}^2_{\rm SN}$ in our best-fit analysis.

\subsection{Baryon Acoustic Oscillations (BAO)}

 From the distribution of SDSS luminous red galaxies, the quantity $A$ via the measurement of the BAO peak is defined as \cite{Eisenstein:2005su}
 \begin{equation}
 A\equiv \Omega_{m0}^{1/2} E(z_{b})^{-1/3}\left({1\over z_{b}}\int_0^{z_{b}}{dz'\over E(z')}\right)^{2/3}~,
 \end{equation}
at the redshift $z_{b}=0.35$. The SDSS BAO measurement  \cite{Eisenstein:2005su} gives $A_{\rm obs}=0.469\ (n_s/0.98)^{-0.35}\pm0.017$, where the scalar spectral
index is taken to be $n_s=0.968$ from the WMAP7 measurement \cite{Komatsu:2010fb}. The $\chi^2$ of the BAO data is given by:
 \begin{equation}
 \chi^2_{\rm BAO} = \frac{(A-A_{\rm obs})^2}{0.017^2} ~.  \label{chi2bao}
 \end{equation}

\subsection{Cosmic Microwave Background (CMB)}

The shift parameter $R$ is defined by \cite{Bond:1997wr}
 \begin{equation}
 R\equiv \Omega_{m0}^{1/2}\int_0^{z_{*}}{dz'\over E(z')}~,
 \end{equation}
where the redshift of the recombination $z_{*}=1091.3$ WMAP7 \cite{Komatsu:2010fb}. The shift parameter $R$, which relates the angular diameter distance to the last
scattering surface, the comoving size of the sound horizon at $z_{*}$ and the angular scale of the first acoustic peak in CMB power spectrum of temperature, has
been measured to be $1.725\pm0.018$ \cite{Komatsu:2010fb}. The $\chi^{2}$ of the CMB data is given by:
 \begin{equation}
 \chi^2_{\rm CMB} =\frac{(R-1.725)^2}{0.018^2} ~. \label{chi2cmb}
 \end{equation}


\end{document}